# Trajectory construction of Dirac evolution

Peter Holland

Green Templeton College, University of Oxford, Oxford, UK

peter.holland@gtc.ox.ac.uk

## Abstract

We extend our programme of representing the quantum state through exact stand-alone trajectory models to the Dirac equation. We show that the free Dirac equation in the angular coordinate representation is a continuity equation for which the real and imaginary parts of the wavefunction, angular versions of Majorana spinors, define conserved densities. We hence deduce an exact formula for the propagation of the Dirac spinor derived from the self-contained first-order dynamics of two sets of trajectories in 3-space together with a mass-dependent evolution operator. The Lorentz covariance of the trajectory equations is established by invoking the 'relativity of the trajectory label'. We show how these results extend to the inclusion of external potentials. We further show that the angular version of Dirac's equation implies continuity equations for currents with non-negative densities, for which the Dirac current defines the mean flow. This provides an alternative trajectory construction of free evolution. Finally, we examine the polar representation of the Dirac equation, which also implies a non-negative conserved density but does not map into a stand-alone trajectory theory. It reveals how the quantum potential is tacit in the Dirac equation.

## 1 Introduction

The wavefunction $\psi$ and its Hilbert space habitat have enjoyed a preeminence in characterizing the quantum state that is deserved only by virtue of the dearth of alternatives. It has been established, however, that this conception of state may be replaced by an exact stand-alone trajectory model, as first shown by the author [1,2]. That is, the burden of quantum evolution may be attributed solely to a congruence of trajectories in the configuration space where $\psi$ evolves. The congruence is, moreover, computed independently of the wavefunction (subject to concordance of the initial conditions in the two pictures). The time-dependent wavefunction thereby becomes a derived entity.

The trajectory theory of state stems in part from the de Broglie-Bohm causal interpretation [3] but it is important to appreciate that it is independent of that view (in particular, it does not require that one of the paths supports a material corpuscle, a notion that requires independent justification [4]), or indeed of any interpretation. In fact, the primary impulse for the trajectory theory has to do with a scarcely remarked lacuna in the history of field theory. In its inchoate 18[th] century elaboration, devoted to the theory of continua, two complementary pictures of a field description emerged [5]. In the material, or Lagrangian, picture the state of a continuous system comprises the displacement function of a continuum of interacting points ('particles') together with an initial density, and its temporal evolution is chronicled by spacetime trajectories. In the spatial, or Eulerian, picture the state is defined in terms of a few key spacetime functions (including density and velocity) and its temporal evolution is recorded at fixed space points. The two pictures are connected by a well-defined mapping (which interchanges dependent and independent variables) but pertain to different physical questions. Insofar as one picture is self-contained, that is, expressible just in terms of its respective notion of state, the other picture may be regarded as 'derived' from it.

Yet, as field theory developed and permeated other disciplines, the eventual lot of the material picture was that it hardly strayed beyond the confines of continuum physics. Post-aether electromagnetic theory, general relativity and quantum theory have all been couched exclusively in spatial terms. There seems no good conceptual or mathematical reason for this lop-sided development of the field approach. Given the potential value of, say, a trajectory formulation of the quantum state, or novel methods of solving field equations, the author initiated a programme to develop material field pictures, for both quantum and non-quantum theories [1,2, 6-15].

The possibility of building a self-contained and exact trajectory counterpart to a spatial-picture field theory depends on how the latter is formulated. We are concerned with spatial theories for which a continuity equation plays a central role, for this is instrumental in establishing a trajectory construction of field propagation (following the continuum physics template). For theories that may be represented using a complex wavefunction $\psi$, two methods of trajectory construction have been developed by the author, characterized by two ways of representing $\psi$ in terms of real functions:

A. Polar variables: $\psi = |\psi|e^{iS/\hbar}$. This is appropriate for the class of theories that may be expressed in the form of the Schrödinger equation (first order in time, quadratic kinetic energy operator) in an $N$-dimensional Riemannian configuration space [2,9]. The time-dependent amplitude $\psi$ is built from a single congruence (i.e., configuration points are linked by at most one path), and the material and spatial pictures are connected by a canonical transformation [1,6]. This class exhibits a generic role for the quantum potential and embraces trajectory-state models for the many-body Schrödinger equation [1,2,9] (which thereby acquires a representation in 3-space [14]), non-relativistic spin $\frac{1}{2}$ systems and quantum fields [9], the mass-zero spin $\frac{1}{2}$ Weyl equation [7], and Maxwell's equations [2] (for further details see [6,11,13]). This approach is set in a wider conceptual and historical context in [16].

B. Real and imaginary parts: $\psi = \psi_R + i\psi_I$. This pertains particularly to field equations that may themselves be expressed in the form of continuity equations, with linear combinations of field variables or their derivatives representing conserved densities. The amplitude $\psi$ is built from two congruences; in a hydrodynamic analogy, the systems are modelled as two interpenetrating, miscible fluids. This method encompasses the Schrödinger equation [8], the massless wave equation [10,15] and the Klein-Gordon equation (which cannot be expressed in the form A) [15].

In both methods the trajectories may obey first- or second-order (in time) equations and their dynamics are driven by inter-trajectory interactions and external potentials. The significance of the paths depends on the context but, whatever property is represented by the density $\rho(x)$ in the various applications of the theory (such as probability or energy), a common element is that the allied paths conserve the quantity $\rho(x)\sqrt{g}\,d^N x$ (see Sect. 2). In method B, $\rho$ may be of either sign. In all cases the material state comprises not just the trajectories but also the associated density at an instant, which we take as $t = 0$.

It is emphasized that it is the trajectory account of *propagation* that dispenses with the wavefunction in these approaches. The *computation* of the trajectories depends intimately on the initial wavefunction $\psi_0 = |\psi_0|e^{iS_0/\hbar}$. For example, for a one-body system treated according to method A, the trajectory-state initial data is derived from $|\psi_0|^2$, which defines the initial density that enters explicitly in the second-order dynamical equation, and the vector $\partial_i S_0/m$, which fixes the initial velocity [1,6][1]. At any other instant, the wavefunction version of state corresponding to the prescribed $\psi_0$ may then be extracted from the trajectory one using a general formula.

An unsolved challenge is to extend the constructive trajectory notion of state to the Dirac equation

$$i\hbar \frac{\partial \Psi^a}{\partial t} = (-i\hbar c(\gamma^0 \gamma^i)^a{}_b \partial_i + mc^2 \gamma^{0a}{}_b)\Psi^b, \quad i,j,\ldots = 1,2,3; \quad a,b,\ldots = 1,2,3,4. \qquad (1)$$

This case poses a problem which afflicts all theories involving discrete indices [3]: to find an appropriate configuration space to set up a correspondence between the wave equation and a continuous trajectory model. This problem has been solved for the analogous Maxwell [2] and Weyl [7] equations. Translating that analysis into the 4-spinor language, the difficulty arises because the wavefunction is defined on the 4-dimensional configuration space whose points are labelled by the coordinates $(x^i, a)$, where the fourth label has just four values: $a = 1,2,3,4$. An apparently natural way to connect the spinor formalism with a continuous trajectory description is to recast it in terms of functions defined on the reduced configuration 3-space labelled by $x^i$, by summing $(x^i, a)$-dependent quantities over the spin indices. For example, using the Dirac 4-current $(j^0, j^i) = c(\Psi^\dagger \Psi, \Psi^\dagger \gamma^0 \gamma^i \Psi)$, trajectories may be connected with the spinor field via the agency of the integral curves of the 3-vector $j^i/j^0$, which is the unique flow that conserves the probability $\Psi^\dagger \Psi d^3 x$ [18,19]. However, it is evident that the 4-curremt, and hence the congruence, carries insufficient information to represent the 8 real components of the spinor field. It is well known [20] that to fully represent a 4-spinor (up to a sign) using (3+1)-tensor fields requires the set of 16 bilinear covariants $\overline{\Psi}\gamma^A \Psi$, $A = 1,\ldots,16$ (subject to 9 identities; for the algebraic construction see [21]), together with a differential quantity (such as the convection current) to represent the (derivative of) the mean phase. The Dirac equation may then be expressed in terms of a (rather complicated) self-contained set of algebraic and differential 3-space tensor relations. The problem with this approach in relation to our programme is that the fields introduced to specify the quantum state that augment the 4-current have no intrinsic connection with, or interpretation in, a trajectory model, that is, they do not map into conserved-density or velocity fields. Thus, in this approach, quantum evolution cannot be attributed solely to an ensemble of trajectories.

---

[1] Note that the author's theory A [1,2,9] has been reproduced under the misleading moniker 'quantum mechanics without wavefunctions' [17].

Instead of reducing the configuration space to $x^i$ and multiplying the number of state functions using the bilinears, we may consider working with the individual spinor components but append to $x^i$ a continuous independent coordinate $s$. Then it transpires that (1) is equivalent to a continuity-type equation in the space $(x^i, s)$ obeyed by the functions $\widetilde{\Psi}^a(x,s) = \Psi^a(x)e^{imc^2s/\hbar}$. We may then seek to emulate our treatment of the non-relativistic Schrödinger equation [8] in method B by treating the real and imaginary parts of each component $\widetilde{\Psi}^a$ as densities (8 in all) and identifying suitable velocities as functions of them. Doing this, (1) becomes a self-contained set of 8 coupled continuity equations. Such a scheme admits a self-contained set of first-order coupled trajectory equations from which $\Psi^a(t)$ may be computed, given the initial $\Psi_0^a$ (see Sect. 2). But, whilst this procedure solves our problem, it is purely formal; the additional dimension $s$ has no connection with the rest of the theory, and the representation of a spinor as a unified geometric object is lost. An alternative approach, that also works with the individual spinor components and uses an enhanced space, is to express the solutions to (1) in terms of solutions to the Klein-Gordon equation and then use our earlier trajectory construction of the latter [15] to build $\Psi$. Thus, the 4-spinor $\Psi = (\gamma^0\,\partial/c\partial t + \gamma^i\partial_i + (mc/i\hbar)I)\mathrm{X}$ solves (1) when the 4-spinor potential $\mathrm{X}^a$ obeys the Klein-Gordon equation $(\Box + (m^2c^2/\hbar^2))\mathrm{X}^a = 0$ [22]. Then, applying our construction [15] to each component $\mathrm{X}^a, a = 1,2,3,4$ (given $\mathrm{X}_0^a$), we may infer $\Psi^a$ as a function of these sets of Klein-Gordon paths. However, this constructive solution is also deficient, although in a different way, for we are seeking a *direct* trajectory construction of Dirac solutions, that is, a trajectory law that maps $\Psi_0^a$ into $\Psi^a(t)$.

The origin of the problems with the candidate constructive-trajectory methods we have outlined (based on summing over spin indices or using individual spinor components) is that they start from the 4-spinor representation of the spin $\frac{1}{2}$ theory. As we showed previously [2,3,7,9], the remedy is to change the representation, from the discrete index $a$ to continuous angle indices $\alpha^r, r = 1,2,3$. As we shall see, the Dirac equation then becomes 'fully differential' in the enhanced configuration space $(x^i, \alpha^r)$, with the $\gamma$ matrices replaced by differential operators and the wavefunction represented by a small number of independent state functions (densities, and velocities defined in terms of them). There is no need for a contrived additional dimension, and the integrity of the wavefunction as a geometric object is maintained. This formulation has a clear connection with a trajectory model defined in the space $(x^i, \alpha^r)$.

Within the framework of the continuous representation we will examine the possibilities for formulating Dirac trajectory constructions corresponding to both methods A and B. Regarding A, an important conclusion of our investigation is that a special property of the Dirac Hamiltonian precludes employing the polar decomposition to connect the field equation to a self-contained single-congruence model, as was possible in the Maxwell and Weyl cases. We show this (in Sect.7.2) as a consequence of the alternative approach that is the focus of the paper. This stems from an apparently unnoticed property of the Dirac equation: that, in the continuous representation, it is a real continuity equation in the space $(x^i, \alpha^r)$ with a complex solution $\psi(x^i, \alpha^r)$. This places the theory within the orbit of method B.

In Sect. 2 we review the connection between the material and spatial pictures of local conservation and determine conditions under which a self-contained first-order law governing trajectory evolution may be used to solve the continuity equation. Since it is rarely used, we present in Sect. 3 the Dirac equation in its angular coordinate formulation, and show in Sect. 4 that it has the form of a continuity equation associated with a brace of real conserved flows. The latter correspond to densities given by the real and imaginary parts of the wavefunction, which are angular representations of Majorana spinors. The preceding results are combined in Sect. 5 to give the evolution equations for the material version of the quantum state. This leads to a fundamental result of the paper, formula (50), which shows that Dirac spinor propagation is generated by two stand-alone Majorana congruences in 3-space together with a mass-dependent evolution operator. The Lorentz covariance of the material picture is established in Sect. 6 by applying the 'relativity of the trajectory label', a concept that was introduced previously in a simplified context [10]. This fills a gap in our analogous work on the Maxwell and Weyl equations [2,7] where the material symmetry that accompanies the spatial Lorentz symmetry was left unspecified. An essential element is the appearance of a novel relativistic transformation rule for velocity, and we go into this in considerable detail with explicit construction of the label transformation functions. In Sect. 7, we bring out a further apparently overlooked aspect of the angular representation: that the Dirac equation implies continuity equations for currents with non-negative densities that do not involve sums over the spin freedom and hence are more detailed than the Dirac current. This observation leads to an alternative trajectory construction, and explains how the quantum potential is tacit in the Dirac equation. In Sect. 8 we examine how our results are modified in the presence of external potentials.

## 2 Trajectory solution of the continuity equation

Consider an $N$-dimensional Riemannian manifold $\mathfrak{M}$ equipped with generalized coordinates $x^\mu$ and metric $g_{\mu\nu}(x)$ where $\mu, \nu, \ldots = 1, \ldots, N$ [2,9,]. Define $g = |\det g_{\mu\nu}|$. The metric is assumed to be independent of the evolution parameter $t$. In the trajectory picture, the state of a continuous system embedded in the space is encoded in the displacement $q^\mu(q_0, t)$ of a point at time $t$, the paths in an ensemble being distinguished by the position $q_0^\mu$ at $t = 0$. We assume that the mapping $q_0^\mu \to q^\mu$ is single-valued and differentiable with respect to $q_0^\mu$ and $t$ to whatever order is necessary, and that the inverse $q_0^\mu(q, t)$ exists and has the same properties. Derivations with respect to the current and initial coordinates are connected by the formula

$$\frac{\partial}{\partial q^\mu} = J^{-1} J_\mu^\nu \frac{\partial}{\partial q_0^\nu} \tag{2}$$

where $J^\nu_\mu$ is the adjoint of the deformation matrix $\partial q^\mu/\partial q^\nu_0$ with

$$\frac{\partial q^\mu}{\partial q^\nu_0} J^\sigma_\mu = J\delta^\sigma_\nu, \quad J^\nu_\mu = \frac{\partial J}{\partial(\partial q^\mu/\partial q^\nu_0)} \tag{3}$$

and the determinant

$$J = \frac{1}{N!}\varepsilon_{\mu_1\ldots\mu_N}\varepsilon^{\nu_1\ldots\nu_N}\frac{\partial q^{\mu_1}}{\partial q^{\nu_1}_0}\ldots\frac{\partial q^{\mu_N}}{\partial q^{\nu_N}_0}, \quad 0 < J < \infty. \tag{4}$$

where $\varepsilon_{\mu_1\ldots\mu_N}$ is the completely antisymmetric symbol. The following two useful formulas follow from (2)-(4), the second being proved using the first:

$$\frac{\partial J^\nu_\mu}{\partial q^\nu_0} = 0, \quad \frac{\partial}{\partial q^\nu_0}\frac{\partial q^\nu_0}{\partial q^\mu} = J\frac{\partial J^{-1}}{\partial q^\mu}. \tag{5}$$

To complete the specification of the state, we need to identify a continuously distributed attribute of the system (the 'charge') that is propagated and conserved by the congruence. Let $\rho_0(q_0)\sqrt{g(q_0)}$ be the initial charge density in $\mathfrak{M}$ ($\rho_0$ may be of either sign). Then the charge in an elementary volume $d^N q_0$ attached to the point $q^\mu_0$ is given by $\rho_0(q_0)\sqrt{g(q_0)}d^N q_0$. Its conservation in the course of the motion is expressed through the relation

$$\rho(q(q_0,t),t)\sqrt{g(q(q_0,t))}d^N q(q_0,t) = \rho_0(q_0)\sqrt{g(q_0)}d^N q_0. \tag{6}$$

This relation is the solution to the differential conservation law

$$\frac{\partial}{\partial t}\left[\rho(q(q_0,t),t)\sqrt{g(q(q_0,t))}d^N q(q_0,t)\right] = 0. \tag{7}$$

(In the material picture, $\partial/\partial t$ is calculated for constant $q^\mu_0$.) The local conservation condition (6) supplies a formula for the density $\rho\sqrt{g}$ at the point $q^\mu$ at time $t$, generated from the initial density by the trajectory linking $q^\mu_0$ to $q^\mu$, in terms of the Jacobian of the transformation between the two sets of coordinates:

$$\rho(q(q,t),t)\sqrt{g(q(q_0,t))} = \sqrt{g(q_0)}J^{-1}(q_0,t)\rho_0(q_0). \tag{8}$$

Hence, if the charge density is the only function of physical interest, *the material state is specified completely by $q^\mu(q_0,t)$ and $\rho_0(q_0)$*.

To translate the local material conservation equation (7) into the spatial picture, the displacement function is made an independent variable: $q^\mu(q_0,t) \to x^\mu$. The density and velocity fields in the two pictures are then connected by the relations

$$\rho(x,t)\sqrt{g(x)} = J^{-1}(q_0,t)\rho_0(q_0)\sqrt{g(q_0)}|_{q_0(x,t)} \tag{9}$$

$$v^\mu(x,t) = \dot{q}^\mu(q_0,t)|_{q_0(x,t)} \tag{10}$$

where $\dot{q}^\mu = \partial q^\mu/\partial t|_{q_0}$. (In the spatial picture, $\partial/\partial t$ is calculated for constant $x^\mu$.) Differentiating (9) with respect to $t$, using the relations $\partial/\partial t|_x = \partial/\partial t|_{q_0} - \dot{q}^\mu\partial/\partial q^\mu$ and $\partial \log J/\partial t = \partial \dot{q}^\mu / \partial q^\mu$ on the right-hand side, and applying (10), we deduce the continuity equation in the spatial picture:

$$\frac{\partial \rho \sqrt{g}}{\partial t} + \partial_\mu (\rho \sqrt{g} v^\mu) = 0 \qquad (11)$$

with $\partial_\mu = \partial/\partial x^\mu$. If the fields vanish at the boundary of $\mathfrak{M}$ this relation implies that $\int \rho(x,t) \sqrt{g(x)}\, d^N x$ is conserved but we shall not need this result (we will be concerned only with integrals over a subset of the independent variables).

The functions $\rho(x,t)$ and $v^\mu(x,t)$ define the state of the system in the spatial picture, and formulas (9) and (10) give the general solution of (11) in terms of the material state, i.e., $q^\mu(q_0,t)$ and $\rho_0(q_0)$. In order to implement these formulas, we need a way to calculate the trajectories by some method that does not require first knowing the spatial density and velocity fields for all $t$. In continuum physics this is achieved by postulating an Euler-type force law (by which we mean the force depends on $\rho$ and $v^\mu$), which in combination with the continuity equation results in a set of self-contained coupled equations for $\rho$ and $v^\mu$. Substituting for these functions in the force law using (9) and (10) then results in a self-contained second-order (in time) equation for the trajectories (given the initial data $\rho_0$ and $\dot{q}_0^\mu$) [23], from whose solutions $\rho(t)$ and $v^\mu(t)$ may be calculated via (9) and (10). This is the method of constructing spatial solutions that we developed in connection with the Schrödinger equation (method A in Sect. 1).

Suppose, however, that the velocity acquires its dependence on $x^\mu, t$ solely via the density function $\rho(x,t)\sqrt{g}$, the known function $\sqrt{g}(x,t)$, and their derivatives: $v^\mu(x,t) = v^\mu(\sqrt{g}, \rho\sqrt{g}, \partial(\rho\sqrt{g}), \dots)$. Then the continuity equation (11) becomes a differential equation just for $\rho$, which is now the sole independent state function. To solve (11) using the paths, we insert (9) in $v^\mu$ so that (10) becomes a self-contained first-order (in time) equation to determine the trajectories given the initial data $\rho_0$:

$$\frac{\partial q^\mu(q_0,t)}{\partial t} = v^\mu(\sqrt{g}, \rho\sqrt{g}, \partial(\rho\sqrt{g}), \dots)|_{x=q(q_0,t),\ \ \rho\sqrt{g}=J^{-1}\rho_0\sqrt{g_0}}. \qquad (12)$$

Here, differentiation with respect to the current coordinates is given by (2). Hence, in the case of a $\rho$-dependent velcoity field, the solution $\rho(x,t)$ may be deduced from the solution to the stand-alone trajectory equation (12) by substituting $q^\mu(q_0,t)$ in (9). As regards the trajectory dynamics, a separate second-order Euler-type force law is superfluous. This trajectory construction corresponds to method B in Sect. 1. Notice that $\rho_0$ enters into the trajectory law.

We shall show that the material version of the Dirac equation comprises two sets of equations (9) and (12), corresponding to a spatial state specified by two densities.

### 3 Fully differential formulation of the Dirac equation

Angular coordinate versions of Dirac's equation have been described previously [24,25]. Here we derive the version given by Dahl [25] but using a different method, and our definitions and conventions are drawn from [3]. We specialize to the six-dimensional manifold $\mathfrak{M} = \mathbb{R}^3 \otimes SU(2)$ with coordinates $x^\mu = (x^i, \alpha^r)$ where $i,j,\dots$ and $r,s,\dots =$ 1,2,3, and $\alpha^r = (\alpha, \beta, \gamma)$ are Euler angles with $\alpha \in [0, \pi]$, $\beta \in [0, 2\pi]$, $\gamma \in [0, 4\pi]$. In the angular coordinate representation, the quantized components of the angular momentum with respect to the space and body axes are, respectively,

$$\widehat{M}_i = -i\hbar A_i^r(\alpha)\partial_r, \quad \widehat{N}_i = -i\hbar B_i^r(\alpha)\partial_r \qquad (13)$$

where $\partial_r = \partial/\partial\alpha^r$ and

$$\left.\begin{aligned} A_i^r &= \begin{pmatrix} -\cos\beta & \sin\beta\cot\alpha & -\sin\beta\csc\alpha \\ \sin\beta & \cos\beta\cot\alpha & -\cos\beta\csc\alpha \\ 0 & -1 & 0 \end{pmatrix} \\ B_i^r &= \begin{pmatrix} -\cos\gamma & -\sin\gamma\csc\alpha & \sin\gamma\cot\alpha \\ -\sin\gamma & \cos\gamma\csc\alpha & -\cos\gamma\cot\alpha \\ 0 & 0 & -1 \end{pmatrix}. \end{aligned}\right\} \qquad (14)$$

It is easily shown that

$$\widehat{N}_i = R_{ij}\widehat{M}_j \qquad (15)$$

where $R_{ij} = B_i^r A_j^{-1r}$ is the 3-rotation matrix written in terms of the Euler angles. The operators $\widehat{M}_i$ and $\widehat{N}_i$ obey the 'ordinary' and 'anomalous' commutation relations, respectively, and commute:[2]

$$[\widehat{M}_i, \widehat{M}_j] = i\hbar\varepsilon_{ijk}\widehat{M}_k, \qquad [\widehat{N}_i, \widehat{N}_j] = -i\hbar\varepsilon_{ijk}\widehat{N}_k, \qquad [\widehat{M}_i, \widehat{N}_j] = 0, \qquad (16)$$

for all $i, j = 1,2,3$ where $\varepsilon_{ijk}$ is the completely antisymmetric symbol. The matrices (14) and $R_{ij}$ obey the following differential identities that will be useful later:

$$\partial_r(\sin\alpha A_i^r) = \partial_r(\sin\alpha B_i^r) = 0 \qquad (17)$$

$$\widehat{M}_i R_{lj} = i\hbar\varepsilon_{ijk}R_{lk}. \qquad (18)$$

The latter relation is proved by substituting (15) into the last relation in (16) and employing the first relation in (16) to derive a linear relation involving the $\widehat{M}_i$s whose coefficients must vanish due to linear independence.

To obtain the angular dependence of the wavefunction, we seek the simultaneous eigenfunctions of $\widehat{M}_i^2, \widehat{M}_3$ and $\widehat{N}_3$. The $s = \frac{1}{2}$ subspace is spanned by the following four basis functions $u_a(\alpha), a = 1,2,3,4$:

$$\left.\begin{aligned} u_1 &= (2\sqrt{2}\pi)^{-1}\cos(\alpha/2)e^{-i(\beta+\gamma)/2}, & u_2 &= -i(2\sqrt{2}\pi)^{-1}\sin(\alpha/2)e^{i(\beta-\gamma)/2}, \\ u_3 &= -i(2\sqrt{2}\pi)^{-1}\sin(\alpha/2)e^{i(\gamma-\beta)/2}, & u_4 &= (2\sqrt{2}\pi)^{-1}\cos(\alpha/2)e^{i(\beta+\gamma)/2}. \end{aligned}\right\} \qquad (19)$$

These obey the orthonormality conditions

$$\int u_a^*(\alpha)\, u_b(\alpha)d\Omega = \delta_{ab}, \quad d\Omega = \sin\alpha\, d\alpha d\beta d\gamma. \qquad (20)$$

A general $s = \frac{1}{2}$ state in the Hilbert space of functions on $\mathfrak{M}$ is then

---

[2] Ref. [24] uses two sets of ordinary angular momentum operators and two independent sets of Euler angles. Dahl's approach uses ordinary and anomalous operators and requires only one set of angles [25].

$$\psi(x,\alpha,t) = \Psi^a(x,t)u_a(\alpha), \quad a = 1,2,3,4, \tag{21}$$

where the coefficients $\Psi^a$ form a Dirac 4-spinor field. The inverse relation is

$$\Psi^a(x,t) = \int u_a^*(\alpha)\,\psi(x,\alpha,t)d\Omega. \tag{22}$$

Evidently, the basis functions effect a transformation between discrete (index $a$) and continuous (indices $\alpha^r$) representations of the wavefunction. Applied to the function (21), the two sets of angular momentum operators obey the defining relations of a Clifford algebra, in addition to (16):

$$\widehat{M}_i\widehat{M}_j + \widehat{M}_j\widehat{M}_i = 2(\hbar/2)^2\delta_{ij}, \quad \widehat{N}_i\widehat{N}_j + \widehat{N}_j\widehat{N}_i = 2(\hbar/2)^2\delta_{ij}. \tag{23}$$

The anticommutation relations are thereby realized by ordinary differential operators.

Having introduced an angular representation for the wavefunction, we next establish the relation between the angular momentum operators and the $\gamma$ matrices by connecting their respective actions on the basis functions. Using (13) and (19), we have

$$\widehat{N}_3 u_a = (\hbar/2)u_b \gamma^{0b}{}_a \gamma^{0b}{}_a, \quad -i\widehat{N}_2\widehat{M}_i u_a = (\hbar/2)^2 u_b \gamma^{ib}{}_a, \quad a,b = 1,2,3,4, \tag{24}$$

where the $\gamma$ matrices have the Dirac representation

$$\gamma^{0a}{}_b = \begin{pmatrix} I & 0 \\ 0 & -I \end{pmatrix}, \quad \gamma^{ia}{}_b = \begin{pmatrix} 0 & \sigma^i \\ -\sigma^i & 0 \end{pmatrix} \tag{25}$$

with the Pauli matrices given by

$$\sigma^1 = \begin{pmatrix} 0 & 1 \\ 1 & 0 \end{pmatrix}, \quad \sigma^2 = \begin{pmatrix} 0 & -i \\ i & 0 \end{pmatrix}, \quad \sigma^3 = \begin{pmatrix} 1 & 0 \\ 0 & -1 \end{pmatrix}. \tag{26}$$

Thus, the $\gamma$ matrices are obtained as matrix representations with respect to the basis functions of certain combinations of angular momentum operators:

$$\int u_a^*((\hbar/2)\widehat{N}_3, -i\widehat{N}_2\widehat{M}_i)\,u_b d\Omega = (\hbar/2)^2(\gamma^0, \gamma^i)^a{}_b \tag{27}$$

In this formalism summation over the spin index $a$ is replaced by differentiation (as in (24)) and/or integration with respect to $\alpha^r$.[3] Thus, the components of the Dirac current $(j^0, j^i)$ have the alternate expressions

$$j^0/c = \Psi^{a*}(x)\Psi^a(x) = \int |\psi(x,\alpha)|^2 d\Omega \tag{28}$$

$$j^i/c = \Psi^{a*}(x)(\gamma^0\gamma^i)^a{}_b\Psi^b(x) = (2/\hbar)^2 \int \psi^*(x,\alpha)\widehat{N}_1\widehat{M}_i\psi(x,\alpha)d\Omega. \tag{29}$$

Using (16) and (27), the differential operator corresponding to $(\hbar/2)^2\gamma^0\gamma^i$ is $\widehat{N}_1\widehat{M}_i$. The Dirac equation (1) may therefore be written in fully differential form as

$$i\hbar \frac{\partial \psi}{\partial t} = -(4ic/\hbar)\hat{N}_1 \widehat{M}_i \partial_i \psi + (2mc^2/\hbar)\hat{N}_3 \psi \tag{30}$$

with initial data $\psi_0(x,\alpha) = \Psi_0^a(x)u_a(\alpha)$. This is the form given by Dahl [25]. Multiplying (30) by $u_a^*$ and applying (20) returns (1). In the guise (30) the Dirac equation appears to be a third-order differential equation for the function $\psi(x,\alpha)$ but, using (15), (16) and (23), we see that it is actually second order (as with the analogous spin 1 Maxwell Hamiltonian [2]). Thus, when applied to a spin $\frac{1}{2}$ function (21), the product of angular momentum operators in (30) is a sum of zeroth and first orders,

$$\hat{N}_1 \widehat{M}_i = \left(\frac{\hbar}{2}\right)^2 R_{1i} - \frac{1}{2} i\hbar \varepsilon_{ijk} R_{1j} \widehat{M}_k, \tag{31}$$

where $R_{1i}$ is a unit vector. Of course, we only consider solutions of (30) for which the $\alpha^r$ dependence is fixed by the basis functions (19).

---

[3] Strictly speaking, $\widehat{M}_i$ and $\widehat{N}_i$ each carry three indices ($\widehat{M}_i(\alpha,\alpha') = -i\hbar A_i^r(\alpha)\partial_r\delta(\alpha-\alpha')$ etc.) so all operations involve differentiation and integration.

Using the results $u_1^* = u_4$, $u_2^* = -u_3$ from (19), and referring to (25), the complex conjugate solution may be written $\psi^* = iu_a \gamma^{2a}{}_b \Psi^{b*}$. The spinor equivalent of complex conjugation is thus charge conjugation. Writing $\psi = \psi_R + i\psi_I$, the real and imaginary parts of the angular wavefunction are connected to the spinor $\Psi^a$ via the relations

$$\left.\begin{array}{ll}\psi_R = \Phi_R^a u_a, & \Phi_R^a = (1/2)(\Psi^a + i\gamma^{2a}{}_b \Psi^{b*}) \\ \psi_I = \Phi_I^a u_a, & \Phi_I^a = (1/2i)(\Psi^a - i\gamma^{2a}{}_b \Psi^{b*}).\end{array}\right\} \tag{32}$$

We see that the discrete correspondents $\Phi_R^a, i\Phi_I^a$ of $\psi$'s components $\psi_R, i\psi_I$ are Majorana spinors (eigenspinors of the charge conjugation operator: $i\gamma^2 \Omega^* = \pm\Omega$ [26]). The superposition $\psi = \psi_R + i\psi_I$ is therefore the angular representative of the decomposition of a Dirac spinor into Majorana spinors: $\Psi^a = \Phi_R^a + i\Phi_I^a$.

A key property of the field equation (30) is that $\psi_R$ and $\psi_I$ are independent solutions. Correspondingly, the functions $\Phi_R$ and $\Phi_I$ satisfy the discrete version (1).

## 4 The Dirac equation as a continuity equation

The metric on $\mathfrak{M} = \mathbb{R}^3 \otimes SU(2)$ is given by [27]

$$g_{\mu\nu} = \begin{pmatrix} \delta_{ij} & 0 \\ 0 & g_{rs} \end{pmatrix}, \quad g_{rs} = \begin{pmatrix} 1 & 0 & 0 \\ 0 & 1 & \cos\alpha \\ 0 & \cos\alpha & 1 \end{pmatrix} \tag{33}$$

with $g = \sin^2\alpha$. The continuity equation (11) in this space then becomes

$$\frac{\partial \sin\alpha\, \rho}{\partial t} + \partial_i(\sin\alpha\, \rho v^i) + \partial_r(\sin\alpha\, \rho v^r) = 0. \tag{34}$$

Here, $v^i(x,\alpha,t)$ and $v^r(x,\alpha,t)$ are translational and angular velocity fields, respectively. An alternative representation of the latter is the angular velocity vector $\omega^i = A_i^{-1r} v^r$ where $A_i^{-1r} A_i^s = \delta^{rs}$.

As in Sect. 2, we introduce trajectories in the configuration space $\mathfrak{M}$ corresponding to the arguments of the density, which are here three translation and three rotation coordinates. Denoting the current and initial trajectory coordinates by $q^\mu(t) = (q^i(t), \theta^r(t))$ and $q_0^\mu = (q_0^i, \theta_0^r)$, respectively, the solution (9) becomes

$$\sin\alpha\,\rho(x,\alpha,t) = J^{-1}(q_0,\theta_0,t)\rho_0(q_0,\theta_0)\sin\theta_0^1|_{\substack{q_0(x,\alpha,t)\\\theta_0(x,\alpha,t)}}. \tag{35}$$

Referring to (2), differentiation with respect to the current cooordinates is given by

$$\frac{\partial}{\partial q^i} = J^{-1}\left(J_i^j\frac{\partial}{\partial q_0^j} + J_i^r\frac{\partial}{\partial \theta_0^r}\right), \qquad \frac{\partial}{\partial \theta^r} = J^{-1}\left(J_r^i\frac{\partial}{\partial q_0^i} + J_r^s\frac{\partial}{\partial \theta_0^s}\right). \tag{36}$$

It is convenient to introduce the dimensionless real operators

$$\widehat{m}_i = 2\widehat{M}_i/i\hbar = -2A_i^r\partial_r, \qquad \widehat{n}_i = 2\widehat{N}_i/i\hbar = -2B_i^r\partial_r, \tag{37}$$

with $\hat{n}_1\widehat{m}_i = -R_{1i} - \varepsilon_{ijk}R_{1j}\widehat{m}_k$ from (31). Then, multiplying by $\sin\alpha$, the Dirac equation (30) becomes

$$\frac{\partial \sin\alpha\,\psi}{\partial t} - \partial_i(c\sin\alpha\,\hat{n}_1\widehat{m}_i\psi) - \partial_{\alpha^3}(2mc^2\sin\alpha\,\psi/\hbar) = 0 \tag{38}$$

where we have used (13) and (14) which give $\hat{n}_3 = 2\partial/\partial\alpha^3$. Comparing (34) with (38), we deduce that the Dirac equation has the form of a continuity equation in $\mathfrak{M}$ with a complex density $\sin\alpha\,\psi$. The functions $\sin\alpha\,\psi_R$ and $\sin\alpha\,\psi_I$ may therefore be interpreted as (real) densities associated with two independent locally conserved flows. For the real part we make the following identifications for the density, translational velocity and angular velocity fields:

$$\rho_R = \psi_R, \quad v_R^i = -\frac{c\hat{n}_1\widehat{m}_i\psi_R}{\psi_R}, \quad v_R^r = -(0,0,\omega) \tag{39}$$

where $\omega = 2mc^2/\hbar$. Similar identifications apply for $\psi_I$.

There are two notable properties of these definitions of the density and translational velocity: these functions are gauge dependent, i.e., they are not invariant under a constant phase shift of $\psi$ (a property exhibited also in the non-relativistic case [8]), and *the speed of translation is bounded from below by the speed of light*. The latter is easily shown using the formula (31), which implies, since $R_{1i}R_{1i} = 1$ and $\boldsymbol{R}_1.(\boldsymbol{R}_1 \times \widehat{\boldsymbol{m}}\psi_R) = 0$, that

$$v_R \equiv \sqrt{v_R^i v_R^i} = c\sqrt{1 + \left(\frac{\boldsymbol{R}_1 \times \widehat{\boldsymbol{m}}\psi_R}{\psi_R}\right)^2} \geq c. \tag{40}$$

The functions $\psi_I$ and $v_I^i$ have the same properties. These features do not signal inconsistencies in the theory; the field equations for $\psi_R$ and $\psi_I$, (38), are collectively globally gauge invariant (which extends to local gauge invariance - see Sect. 8)) and individually Lorentz covariant (Sect. 6). In particular, the angular mean over the two translational flows generates the gauge-invariant, future-causal flow defined by the Dirac current (Sect. 7.1).

## 5 Trajectory construction of a time-dependent Dirac spinor field

We now combine the results of Sections 2-4 to obtain the material version of the Dirac equation, and use its solutions to derive formulas for the propagation of each of the Majorana amplitudes $\psi_R(t)$ and $i\psi_I(t)$ from single congruences. The spatial general solution $\psi(t)$, and hence $\Psi^a(t)$, follows by superposition.

It will be observed from (39) that the dependence of the translational velocity $v_R^i$ on its arguments $x^i, \alpha^r, t$ derives just from the functions $\sin\alpha$ and $\sin\alpha\,\psi_R$ and their derivatives (that is, $v_R^i = -c(\sin\alpha/\sin\alpha\,\psi_R)\hat{n}_1\widehat{m}_i(\sin\alpha\,\psi_R/\sin\alpha)$), while the angular velocity is constant. The velocity fields are therefore of the type appearing on the right-hand side of (12). The corresponding six equations for the coordinates $q_R^i, \theta_R^r$,

$$\frac{\partial q_R^i(q_{R0},\theta_{R0},t)}{\partial t} = v_R^i(x=q_R, \alpha=\theta_R, t) = -\frac{c\hat{n}_1(\alpha)\widehat{m}_i(\alpha)\psi_R(x,\alpha)}{\psi_R(x,\alpha)}\bigg|_{\substack{x=q_R(q_{R0}, \\ \alpha=\theta_R(q_{R0}}} \quad (41)$$

$$\frac{\partial \theta_R^r(q_{R0},\theta_{R0},t)}{\partial t} = v_R^r(x=q_R, \alpha=\theta_R, t) = -\omega\delta_3^r, \quad (42)$$

are thus of the form (12). In (41), following (35), we substitute

$$\psi_R(x=q_R,\alpha=\theta_R,t)\sin\theta_R^1(q_{R0},\theta_{R0}) = J_R^{-1}(q_{R0},\theta_{R0},t)\psi_{R0}(q_{R0},\theta_{R0})\sin\theta_{R0}^1 \quad (43)$$

where $\psi_{R0} = \Phi_{R0}^a(q_{R0})u_a(\theta_{R0})$ from (32). The differential operators $\hat{n}_1$ and $\widehat{m}_i$ are given by (36) and (37).

The set of equations (41)-(43), together with the similar set for $\psi_I$ that employ a second set of trajectories $q_I^i(q_{I0},\theta_{I0},t), \theta_I^r(q_{I0},\theta_{I0},t)$, constitute *the material version of the Dirac equation*. The material counterpart of the spatial quantum state $\psi$ is defined by the two sets of trajectories and the initial functions $\psi_{R0}$ and $\psi_{I0}$. It is evident that the latter prescribed data participate in the trajectory equation (41).

The solution of (42) for the angle coordinates is immediate,

$$\theta_R^1 = \theta_{R0}^1, \quad \theta_R^2 = \theta_{R0}^2, \quad \theta_R^3 = \theta_{R0}^3 - \omega t, \quad (44)$$

and is independent of $\psi_{R0}$ and $q_{R0}$. The angular velocity vector $\omega_R^i = -\omega R_{3i}(\theta_{R0})$ is therefore constant. These results for the angles simplify matters: $J_R$ reduces to the 3-determinant $\det(\partial q_R/\partial q_{R0})$ and $\sin\theta_R^1$ drops out. Then (43) becomes

$$\psi_R(x=q_R,\alpha=\theta_R,t) = J_R^{-1}(q_{R0},\theta_{R0},t)\Phi_{R0}^a(q_{R0})u_a(\theta_{R0}). \quad (45)$$

Inserting this formula in (41) gives a self-contained set of three coupled differential equations to determine the 3-trajectories $q_R^i$. Conversely, once in possession of the so-computed trajectories, the spatial version of the formula (45), which gives the solution to the real part of the Dirac equation (38), follows by inserting $q_{R0}^i(x,\alpha,t), \theta_{R0}^r(\alpha,t)$ on the right-hand side. It is instructive to replace the Jacobian in (45) by an integral and give the solution in propagator form:

$$\psi_R(x,\alpha,t) = \int \delta(x - q_R(q_0,\theta_{R0},t))\delta(\alpha - \theta_R(\theta_{R0},t))\Phi_{R0}^b(q_0)u_a(\theta_0)\,d^3q_0 d\Omega_0 \quad (46)$$

The trajectory construction of the Majorana spinor solution $\Phi_R^a$ now follows from (22) by inversion. Thus, multiplying (46) by $\sin\alpha\, u_a^*(\alpha)$, integrating over $\alpha^r$, and noting from (19) that $u_a^*(\theta_R) = U_b^a u_b^*(\theta_{R0})$ where

$$U_b^a(t) = \cos(\omega t/2)\delta_b^a - i\gamma^{0a}{}_b\sin(\omega t/2), \qquad \omega = 2mc^2/\hbar, \tag{47}$$

is a mass-dependent unitary evolution operator, we get

$$\Phi_R^a(x,t) = U_c^a(t) \int \delta(x - q_R(q_0, \theta_0, t))\, \Phi_{R0}^b(q_0) u_b(\theta_0) u_c^*(\theta_0)\, d^3q_0 d\Omega_0. \tag{48}$$

It is readily confirmed that this function indeed obeys (1) using the formulas $dU_c^a/dt = -(i\omega/2)\gamma^{0a}{}_b U_c^b$, $\dot{q}_R^i\delta = v_R^i\delta$, $\partial\delta/\partial t = -\partial_i(v_R^i\delta)$, $\psi_R v_R^i = c u_a (\gamma^0\gamma^i)^a{}_b \Phi_R^b$, and (20) (we have written $\delta \equiv \delta(x - q_R)$ here).

Repeating this procedure to obtain the imaginary part $\psi_I$ in terms of the second set of trajectories (for which, in particular, $\theta_I^r = \theta_{I0}^r - \delta_3^r\omega t$), the amplitude $\psi$ at each configuration point $(x^i, \alpha^r)$ is generated by (at most) two configuration space paths. Whilst the $R$ and $I$ trajectory equations are solved independently, the condition that both paths arrive at the same point at time $t$, namely,

$$x^i = q_R^i(q_{R0}, \theta_{R0}, t) = q_I^i(q_{I0}, \theta_{I0}, t), \qquad \alpha^r = \theta_R^r(q_{R0}, \theta_{R0}, t) = \theta_I^r(q_{I0}, \theta_{I0}, t), \tag{49}$$

enforces six relations among the initial coordinates: $q_{I0}^i = q_{I0}^i(q_{R0}, \theta_{R0}, t), \theta_{R0}^r = \theta_{I0}^r$.

In sum, we obtain the following exact propagator expression for the time-dependent Dirac spinor as a superposition of Majorana amplitudes ($\Psi^a = \Phi_R^a + i\Phi_I^a$) built from two independent congruences $q_R^i(q_{R0}, \theta_{R0}, t), \theta_R^r(\theta_{R0}, t)$ and $q_I^i(q_{I0}, \theta_{I0}, t), \theta_I^r(\theta_{I0}, t)$:

$$\Psi^a(x,t) = U_c^a(t) \int [\delta(x - q_R(q_0, \theta_0, t))\, \Phi_{R0}^b(q_0) \tag{50}$$
$$+ i\delta(x - q_I(q_0, \theta_0, t))\, \Phi_{I0}^b(q_0)] u_b(\theta_0) u_c^*(\theta_0)\, d^3q_0 d\Omega_0.$$

Here $\Phi_{R0}^b$ and $\Phi_{I0}^b$ represent the prescribed initial state $\Psi_0^a$ via the relations (32). We see that the evolution is driven by two sets of trajectories in 3-space, while the two sets of angular trajectories contribute through the (same) operator $U_b^a$ given in (47).

As a simple example, we compute the time dependence of a spinor whose initial value is $\Psi_0^a = \delta_1^a$, so that $\psi_{R0} = \frac{1}{2}(u_1 + u_4)$ and $\psi_{I0} = \frac{1}{2i}(u_1 - u_4)$. To evaluate the right-hand side of (41), we first insert the angle solution (44) in $\hat{n}_1 \widehat{m}_i$ and, using (13), (14) and (36), obtain

$$\hat{n}_1(\theta_R)\widehat{m}_i(\theta_R) = [\cos(\omega t)\hat{n}_1(\theta_{R0}) + \sin(\omega t)\hat{n}_2(\theta_{R0})]\widehat{m}_i(\theta_{R0}). \tag{51}$$

The solution to (41) is then

$$q_R^i = q_{R0}^i + d^i - (c/\omega)\sec\varphi_+ \left(\sin(\varphi_+ - \omega t), \cos(\varphi_+ - \omega t), -\tan\left(\tfrac{1}{2}\theta_{R0}^1\right)\cos(\varphi_- - \omega t)\right) \tag{52}$$

where $\varphi_\pm = \frac{1}{2}(\theta_{R0}^3 \pm \theta_{R0}^2)$ and $d^i = (c/\omega)\sec\varphi_+(\sin\varphi_+, \cos\varphi_+, -\tan(\theta_{R0}^1/2)\cos\varphi_-)$. For each choice of $\theta_{R0}^r$, the trajectory (52) describes a 'nutation': a fixed-radius circle coplanar with the $x^1 x^2$-plane and a fixed-amplitude oscillation along $x^3$, constrianed by $|\dot{q}_R^i| \geq c$. For this solution, $J_R = 1$. Repeating this analysis for $\psi_I$, we deduce, using (45) and its $\psi_I$ counterpart, that $\Psi^a(x,t) = \delta_1^a e^{-imc^2 t/\hbar}$, which is indeed the solution to the free Dirac equation for a positive-energy plane wave with zero wave vector.

## 6 Lorentz covariance of the congruences

### *6.1 Spatial picture*

Denoting the boost parameter by $\varepsilon^i = u^i/c$, $|\varepsilon^i| \ll 1$, an infinitesimal Lorentz transformation is defined, in the angular language, by $(i, j, \ldots; r, s, \ldots = 1,2,3)$[4]

$$\left.\begin{array}{c} x'^i = x^i - \varepsilon^i ct, \quad t' = t - \varepsilon^i x^i/c, \quad \alpha'^r = \alpha^r, \\ \psi'(x', \alpha, t') = \psi(x, \alpha, t) + \frac{1}{2}\varepsilon^i \widehat{n}_1 \widehat{m}_i \psi(x, \alpha, t), \end{array}\right\} \quad (53)$$

where $\psi'(x', \alpha, t') = \Psi'^a(x', t') u_a(\alpha)$. The form invariance of Dirac's equation (30) means that

$$\frac{\partial \psi'}{\partial t'} - \partial'_i(c\widehat{n}_1 \widehat{m}_i \psi') - \widehat{n}_3(mc^2 \psi'/\hbar) = 0. \quad (54)$$

The invariance of the angular momentum operators in this equation corresponds to the numerical invariance of the $\gamma$ matrices in the discrete formulation; in both cases, a Lorentz transformation on the vector index is 'undone' by a transformation on the two spin indices (cf. footnote 3).

An important aspect of (53) is that the mapping of $\psi$ is real and hence $\psi_R$ and $\psi_I$ transfrorm into themselves. Considering the real part, the transformation laws of the density and velocities (39) implied by (53) are, to first order,

$$\left.\begin{array}{c} \psi'_R = \psi_R(1 - (1/2c)\varepsilon^i v_R^i), \\ v_R'^i = -\dfrac{c\widehat{n}_1 \widehat{m}_i \psi'_R}{\psi'_R} = v_R^i + \dfrac{1}{2c}\varepsilon^j v_R^j v_R^i - \dfrac{c}{2}\varepsilon^j \dfrac{\widehat{n}_1 \widehat{m}_i \widehat{n}_1 \widehat{m}_j \psi_R}{\psi_R}, \\ v_R'^r = v_R^r = -\omega \delta_3^r. \end{array}\right\} \quad (55)$$

These variables therefore transform as a closed set and so the continuity-equation form of Dirac's equation is preserved. In particular, it is obvious that the perpetual superluminality (40) of the flow is preserved under the transformation: $v'_R \geq c$. It is convenient to leave the product $\widehat{n}_1 \widehat{n}_1 (= -1)$ explicit in the formula for $v_R'^i$.

### *6.2 Material picture*

We shall demonstrate that the real part of the Dirac equation in its material incarnation, namely, the set of equations (41)-(43), is Lorentz covariant. That is,

$$\psi'_R = J_R'^{-1} \psi'_{R0}, \qquad v_R'^i = \dot{q}_R'^i, \qquad v_R'^r = \dot{\theta}_R'^r. \quad (56)$$

---

[4] Eq. (30) is covariant under the infinitesimal angle transformation $\alpha'^r = \alpha^r + \eta^i \widehat{n}_1 A_i^r$, which corresponds to the identity symmetry of (1). A symmetry of (30) corresponding to a continuous symmetry of (1) may therefore exhibit this angular freedom.

A first point to note is that the differential transformation law (55) of the translational velocity $v_R^i$ is unfamiliar; it is neither a Lorentz 3-vector (i.e., $\equiv u^i/u^0$ where $(u^0, u^i)$ is a 4-vector) nor obviously part of any other spacetime tensor. On the other hand, if, as we assume, the material variables $q_R^i, t$ transform like the spatial variables $x^i, t$, we might expect that the material velocity $\dot{q}_R^i$ is a Lorentz 3-vector. It appears then that the material picture breaks relativistic covariance in that the law of motion $\dot{q}_R^i = v_R^i$ equates quantities having different transformation properties. In fact, this reasoning is flawed. The origin of the apparent disparity in transformation rules is the assumption, tacit in the usual transformation of a 3-velocity, that a trajectory label (here $(q_{R0}, \theta_{R0})$) is an invariant quantity, i.e., that the same label is attached to the original and transformed paths. But, in the field theory of trajectories we are advancing, the arena of independent variables to which transformations apply is label-time space $(q_{R0}, \theta_{R0}, t)$. Then, a given trajectory will generally be attributed different labels by relatively moving observers (adopting the passive viewpoint). We refer to this feature as 'relativity of the label' [10] or 'relativity of identity'. We will show that the transformation of $\dot{q}_R^i$ contains terms, additional to those of a usual 3-velocity, that represent a possible change in label so as to mirror precisely the transformation of $v_R^i$ given in (55) (see (60) below). This idea is consistent in a continuum setting because, whatever the label needs to be to fulfil the transformation, that value will be available to each observer.

A further important property of the label transformation in demonstrating Lorentz covariance is that it ensures that the set of trajectories may be assigned a common time in each frame.

The material-picture infinitesimal substitution corresponding to (53) is effected in label-time space $(q_{R0}, \theta_{R0}, t)$ and on functions therein as follows:

$$\left.\begin{array}{l} t' = t - \varepsilon^i q_R^i(q_0, \theta_0, t)/c\,, \qquad q_R'^i(q_0', \theta_0', t') = q_R^i(q_0, \theta_0, t) - \varepsilon^i ct, \\ q_{R0}'^i(q_0, \theta_0, t) = q_{R0}^i + \varepsilon^j \xi_j^i(q_0, \theta_0, t), \qquad \theta_{R0}'^r(q_0, \theta_0, t) = \theta_{R0}^r + \varepsilon^j \bar{\xi}_j^r(q_0, \theta_0, t), \\ \theta_R'^r(q_0', \theta_0', t') = \theta_R^r(q_0, \theta_0, t), \quad \psi_{R0}'(q_{R0}', \theta_{R0}') = \psi_{R0}(q_{R0}, \theta_{R0}) + \varepsilon^j X^j(q_{R0}, \theta_{R0}). \end{array}\right\} \quad (57)$$

Our goal is to discover the label and initial-density transformation functions $\xi_j^i$, $\bar{\xi}_j^r$ and $X^j$ that generate a solution of the material version of the Dirac equation (56) in any frame, given the solution $q_R^i, \theta_R^r$ in the original frame.

A notable feature of corresponding symmetries in the spatial and material pictures is that they are not one-to-one, because the identity transformation in the former maps into a time-independent diffeomorphism, a relabelling of the paths, in the latter [10, 11]. This relabelling symmetry is therefore a component of the material symmetry corresponding to any continuous spatial symmetry. It expresses the freedom to choose a label other than the initial position when identifying a trajectory [23]. As we show below, this latitude in the material description can be suppressed by requiring that the condition $q^i(t=0) = q_0^i$, $\theta^r(t=0) = \theta_0^r$ is maintained in all frames.

To proceed, we write down the material version of (55) using the expressions (41)-(43) and (56) for the density and translational and angular velocities:

$$
\left.
\begin{aligned}
&J_R'^{-1}(q_{R0}', \theta_{R0}', t')\psi_{R0}'(q_{R0}', \theta_{R0}') \\
&\quad = J_R^{-1}(q_{R0}, \theta_{R0}, t)\psi_{R0}(q_{R0}, \theta_{R0})\bigl(1 - (1/2c)\varepsilon^i \dot{q}_R^i(q_{R0}, \theta_{R0}, t)\bigr) \quad (a) \\
&\dot{q}_R'^i = \dot{q}_R^i + \frac{1}{2c}\varepsilon^j \dot{q}_R^j \dot{q}_R^i - c\varepsilon^i + \frac{c}{2}\varepsilon^j \frac{\hat{n}_1 \widehat{m}_j \hat{n}_1 \widehat{m}_i (J_R^{-1}\psi_{R0})}{J_R^{-1}\psi_{R0}} \quad (b) \\
&\dot{\theta}_R'^r = \dot{\theta}_R^r = -\omega \delta_3^r. \quad (c)
\end{aligned}
\right\} \quad (58)
$$

In (58b) we have used (23) to reverse the order of the factors $\widehat{m}_i, \widehat{m}_j$ in (55).

Determining $\bar{\xi}_j^r$ is straightforward. From (58c) the solution for the angles in the primed frame is $\theta_R'^r = \theta_{R0}'^r - \delta_3^r \omega t'$ where we take $\theta_R'^r(t'=0) = \theta_{R0}'^r$. Substituting for $t'$ from the first relation in (57), and using the fourth and fifth relations, gives

$$\bar{\xi}_j^r(q_0, \theta_0, t) = -\delta_3^r \omega q_R^j(q_{R0}, \theta_{R0}, t)/c, \tag{59}$$

which is thereby fixed uniquely in terms of given quantities.

To find $\xi_j^i$, we shall use the following expression for $\dot{q}_R'^i$ derived from (57):

$$\frac{\partial q_R'^i}{\partial t'} = \frac{\partial q_R^i}{\partial t} + \varepsilon^j \left(\frac{1}{c}\frac{\partial q_R^j}{\partial t}\frac{\partial q_R^i}{\partial t} - c\delta^{ij} - \frac{\partial \xi_j^k}{\partial t}\frac{\partial q_R^i}{\partial q_{R0}^k} - \frac{\partial \bar{\xi}_j^r}{\partial t}\frac{\partial q_R^i}{\partial \theta_{R0}^r}\right). \tag{60}$$

Here, the first three terms on the right-hand side characterize the usual transformation of a Lorentz 3-velocity and the remaining two terms represent changes in the labels, as mentioned above. Comparing the expressions (58b) and (60) for $\dot{q}_R'^i$ and inserting (59) gives a formula for $\dot{\xi}_j^i$:

$$\frac{\partial \xi_j^k}{\partial t}\frac{\partial q_R^i}{\partial q_{R0}^k} = \frac{1}{2c}\dot{q}_R^j \dot{q}_R^i - \frac{\omega}{c}\dot{q}_R^j \frac{\partial q_R^i}{\partial \theta_{R0}^3} - c\hat{n}_1 \widehat{m}_j \hat{n}_1 \widehat{m}_i \psi_R / 2\psi_R \Big|_{\substack{x=q_R(q_{R0}, \theta_{R0}, t) \\ \alpha=\theta_R(\theta_{R0}, t)}} \tag{61}$$

where we insert $\psi_R = J_R^{-1}\psi_{R0}$. Noting that the inverse of the matrix $\partial q_R^i/\partial q_{R0}^k$ is $\partial q_{R0}^i/\partial q_R^k$ (due to the functional dependence of $\theta_R^r(\theta_{R0}, t)$), we have

$$\xi_j^i(q_{R0}, \theta_{R0}, t) = \int_0^t Y_j^i(q_{R0}, \theta_{R0}, t)dt + \xi_{j0}^i(q_{R0}, \theta_{R0}) \tag{62}$$

where

$$Y_j^i = \frac{\partial q_{R0}^i}{\partial q_R^k}\left(\frac{1}{2c}\dot{q}_R^j \dot{q}_R^k - \frac{\omega}{c}\dot{q}_R^j \frac{\partial q_R^k}{\partial \theta_{R0}^3} - c\frac{\hat{n}_1 \widehat{m}_j \hat{n}_1 \widehat{m}_k \psi_R}{2\psi_R}\right)\Bigg|_{\substack{x=q_R(q_{R0}, \theta_{R0}, t) \\ \alpha=\theta_R(\theta_{R0}, t)}}. \tag{63}$$

As anticipated above, the (initial) time-independent label function $\xi_{j0}^i$ in (62) is fixed by the requirement that $q_{R0}'^i$ is the initial value of $q_R'^i$. In showing this, the following development refers to (57). Setting $t'=0$, the corresponding time in the original frame is given by the solution to $t = \varepsilon^i q_R^i(q_{R0}, \theta_{R0}, t)/c$, which has the form $t = \varepsilon^i f^i(q_{R0}, \theta_{R0})$. Now, to first order, $q_R'^i(t'=0) = q_R^i(t = \varepsilon^j f^j)$ and we require that this is equal to $q_{R0}'^i = q_{R0}^i + \varepsilon^j \xi_j^i(t = \varepsilon^j f^j)$. Then, Taylor expanding $q_R^i(t = \varepsilon^j f^j)$ $= q_{R0}^i + \varepsilon^j f^j \dot{q}_{R0}^i$ and $\xi_j^i(t = \varepsilon^k f^k) = \xi_{j0}^i + \varepsilon^k f^k \partial \xi_{j0}^i/\partial t$, the requisite equality implies

$$\xi_{j0}^i(q_{R0}, \theta_{R0}) = f^j(q_{R0}, \theta_{R0})\dot{q}_{R0}^i(q_{R0}, \theta_{R0}). \tag{64}$$

The label function (62) is therefore determined entirely in terms of given functions.

Finally, we ascertain $\psi'_{R0}$ from (58a). To obtain the transformed determinant $J'_R = \det(\partial q'_R/\partial q'_{R0})$, we use the following relation derived from (57):

$$\frac{\partial q'^i_R}{\partial q'^l_{R0}} = \frac{\partial q^i_R}{\partial q^l_{R0}} + \varepsilon^j \left(\frac{1}{c}\frac{\partial q^j_R}{\partial q^l_{R0}}\frac{\partial q^i_R}{\partial t} - \frac{\partial \xi^k_j}{\partial q^l_{R0}}\frac{\partial q^i_R}{\partial q^k_{R0}} - \frac{\partial \bar{\xi}^r_j}{\partial q^l_{R0}}\frac{\partial q^i_R}{\partial \theta^r_{R0}}\right). \tag{65}$$

This implies

$$J'_R(q'_{R0}, \theta'_{R0}, t') = J_R(q_{R0}, \theta_{R0}, t)\left[1 + \varepsilon^j\left(\frac{1}{c}\dot{q}^j_R - \frac{\partial \xi^k_j}{\partial q^k_{R0}} + \frac{\omega}{c}\frac{\partial q^j_R}{\partial \theta^3_{R0}}\right)\right]. \tag{66}$$

Instead of determining the total variation $X^j$ in (57) directly, it is convenient to use as the unknown function the functional variation of $\psi_{R0}$, i.e., $\varepsilon^i P_i(a) = \psi'_{R0}(a) - \psi_{R0}(a)$, which appears in the Taylor expansion of the transformed initial density,

$$\psi'_{R0}(q'_{R0}, \theta'_{R0}) = \psi_{R0}(q_{R0}, \theta_{R0}) + \varepsilon^j\left(\frac{\partial \psi_{R0}}{\partial q^i_{R0}}\xi^i_j - \frac{\omega}{c}\frac{\partial \psi_{R0}}{\partial \theta^3_{R0}}q^j_R + P_j\right), \tag{67}$$

where we have used (59) ($X^j$ is the term in brackets in (67)). Then, combining (66) and (67), we obtain from (58a)

$$\frac{\partial}{\partial q^i_{R0}}(\psi_{R0}\xi^i_j) + P_j(q_{R0}, \theta_{R0}) = \frac{\omega}{c}\frac{\partial}{\partial \theta^3_{R0}}(\psi_{R0} q^j_R) + \frac{1}{2c}\psi_{R0}\dot{q}^j_R. \tag{68}$$

As a final step, we find $P_j$ by substituting for $\xi^i_j$ from (62). For this purpose, we first evaluate $\partial(\psi_{R0} Y^i_j)/\partial q^i_{R0}$. This is an intricate calculation and we highlight only the significant steps. Using the second formula in (5), (63) implies

$$\frac{\partial}{\partial q^i_{R0}}(\psi_{R0} Y^i_j) = J_R \frac{\partial}{\partial q^k_R}\left[J_R^{-1}\psi_{R0}\left(\frac{1}{2c}\dot{q}^j_R \dot{q}^k_R - \frac{\omega}{c}\dot{q}^j_R \frac{\partial q^k_R}{\partial \theta^3_{R0}}\right)\right. \\ \left. -(c/2)\hat{n}_1 \widehat{m}_j \hat{n}_1 \widehat{m}_k \psi_R\big|_{\substack{x=q_R(q_{R0},\theta_{R0},t) \\ \alpha=\theta_R(\theta_{R0},t)}}\right]. \tag{69}$$

To simplify the right-hand side of (69), we replace the term $\partial_k(c\hat{n}_1\widehat{m}_k\psi_R)$ by $\partial\psi_R/\partial t|_{x,\alpha} - \partial_{\alpha^3}(2mc^2\psi_R/\hbar)$ from Dirac's equation (38); replace $\partial\psi_R/\partial t|_{x,\alpha}$ by $\partial/\partial t|_{q_0,\theta_0} - \dot{q}^i_R\,\partial/\partial q^i_R - \dot{\theta}^r_R\,\partial/\partial \theta^r_R$; and apply the formulas $\partial \log J_R/\partial t = \partial \dot{q}^i_R/\partial q^i_R$ and $\partial/\partial \theta^3_{R0} = \partial/\partial \theta^3_R + (\partial q^i_R/\partial \theta^3_{R0})\,\partial/\partial q^i_R$. The result is

$$\frac{\partial}{\partial q^i_{R0}}(\psi_{R0} Y^i_j) = \frac{\partial}{\partial t}\left(\frac{\omega}{c}\frac{\partial}{\partial \theta^3_{R0}}(\psi_{R0} q^j_R) + \frac{1}{2c}\psi_{R0}\dot{q}^j_R\right). \tag{70}$$

Integrating (70) with respect to $t$ and using (62) gives $\partial(\psi_{R0}\xi^i_j)/\partial q^i_{R0}$. Substituing this in (68), the time-dependent terms drop out and we get

$$P_j(q_{R0}, \theta_{R0}) = \frac{\omega}{c}q^j_{R0}\frac{\partial \psi_{R0}}{\partial \theta^3_{R0}} - \frac{\partial}{\partial q^i_{R0}}(\psi_{R0}\xi^i_{j0}) + \frac{1}{2c}\psi_{R0}\dot{q}^j_{R0}. \tag{71}$$

As a check, this relation coincides with (68) evaluated at $t = 0$.

To summarize, we have established the Lorentz covariance of the material version of the real part of the Dirac equation, i.e., the validity of (56), when the material variables undergo the transformation (57) with the label and initial-density functions given by (59), (62)-(64) and (71). The condition 'label = initial position' is covariant under this transformation. Repeating the procedure for $\psi_I$, and noting from (57) that the constraint (49) is covariant, we have thus demonstrated the Lorentz covariance of the trajectory version of the Dirac equation. 3-space rotational covariance may be demonstrated similarly.

## 7 Novel non-negative conserved densities implied by the Dirac equation

### 7.1 Partial densities

We present here two additional ways of formulating the Dirac equation that feature continuity equations, and examine their propensity for trajectory formulations. In contrast with our method hitherto, where the conserved densities $\sin\alpha\,\psi_R$ and $\sin\alpha\,\psi_I$ are of variable sign, the signature of these alternative versions is that the densities are non-negative.

Eqs. (28) and (29) show that the future-causal flow defined by the Dirac current ($j^0 \geq 0, j^0 j^0 - j^i j^i \geq 0$) may be expressed as the mean over the two superluminal translational flows. Referring to the definition (39) of velocity, we have

$$\Psi^\dagger \Psi = \int (\psi_R{}^2 + \psi_I{}^2) d\Omega \tag{72}$$

$$c\Psi^\dagger \gamma^0 \gamma^i \Psi = \int (\psi_R{}^2 v_R^i + \psi_I{}^2 v_I^i) d\Omega. \tag{73}$$

In these formulas it is the non-negative weights $\sin\alpha\,\psi_R{}^2$ and $\sin\alpha\,\psi_I{}^2$ that are attributed to the translational velocities, rather than the functions $\sin\alpha\,\psi_R$ and $\sin\alpha\,\psi_I$ that are conserved by the flows the velocities represent. Are the functions $\sin\alpha\,\psi_R{}^2$ and $\sin\alpha\,\psi_I{}^2$ also densities obeying continuity equations? From (38) we get, using the identifications of velocities in (39),

$$\partial \psi_R{}^2 / \partial t + \partial_i (\psi_R{}^2 v_R^i) + \partial_r (\psi_R{}^2 v_R^r) = -\psi_R{}^2 \partial_i v_R^i, \tag{74}$$

where we used $\partial_r v_R^r = 0$. Rewriting the right-hand side of (74) using the identity

$$\psi_R{}^2 \partial_i v_R^i = \hat{n}_1 (c\widehat{m}_i \psi_R \partial_i \psi_R) - \widehat{m}_i (c\psi_R \hat{n}_1 \partial_i \psi_R), \tag{75}$$

multiplying through by $\sin\alpha$, and utilizing the identities (17), (74) can be written

$$\partial \sin\alpha\,\psi_R{}^2 / \partial t + \partial_i (\sin\alpha\,\psi_R{}^2 v_R^i) + \partial_r (\sin\alpha\,\psi_R{}^2 \tilde{v}_R^r) = 0 \tag{76}$$

with

$$\tilde{v}_R^r = v_R^r + 2c(A_i^r \psi_R \hat{n}_1 \partial_i \psi_R - B_1^r \widehat{m}_i \psi_R \partial_i \psi_R)/\psi_R{}^2, \tag{77}$$

$v_R^i = -c\hat{n}_1 \widehat{m}_i \psi_R/\psi_R$ and $v_R^r = -\delta_3^r \omega$. Hence, the non-negative function $\sin\alpha\,\psi_R{}^2$ indeed obeys a continuity equation in $\mathfrak{M}$. This involves the same translational velocity that generates the evolution of the density $\sin\alpha\,\psi_R$ and a modified angular velocity. Eq. (76) and the similar equation for $\sin\alpha\,\psi_I{}^2$ collectively provide an alternative version of Dirac's equation. Since the functions of $\psi_R$ in (76) and (77) can be replaced by functions of $\sin\alpha$ and $\sin\alpha\,\psi_R^2$ and their derivatives, we obtain a flow of the type (12). Hence, an alternative trajectory construction of $\psi$ follows for which $\sin\alpha\,\psi_R{}^2 = J_R^{-1}\sin\theta_{R0}^1\,\psi_{R0}{}^2$ and $\sin\alpha\,\psi_I{}^2 = J_I^{-1}\sin\theta_{I0}^1\,\psi_{I0}{}^2$. The signs of the functions $\psi_R$ and $\psi_I$ deduced from these formulas are fixed by the initial data $\psi_0$. We conclude that the time-dependent Dirac spinor may be constructed from two conserved congruences, each of which is associated with a non-negative density.

To obtain the conserved current in the spinor formalism corresponding to $\psi_R{}^2$, we write $\psi_R = \Phi_R^a u_a = \Phi_R^{a*}u_a^*$ and use (20) to get

$$\int \psi_R{}^2 d\Omega = \Phi_R^\dagger \Phi_R = \frac{1}{4}(2\Psi^\dagger \Psi + i(\Psi^\dagger \gamma^2 \Psi^* + \Psi^T \gamma^2 \Psi)) \tag{78}$$

$$\int \psi_R{}^2 v_R^i d\Omega = \Phi_R^\dagger \gamma^0 \gamma^i \Phi_R = \frac{1}{4}\left(2\Psi^\dagger \gamma^0 \gamma^i \Psi + i(\Psi^\dagger \gamma^0 \gamma^i \gamma^2 \Psi^* + \Psi^T \gamma^2 \gamma^0 \gamma^i \Psi)\right). \tag{79}$$

We thus obtain the Majorana current. Since the first terms on the right-hand sides of (78) and (79) together constitute ($\frac{1}{2}$ times) the Dirac 4-current, it follows that the remaining terms jointly form a 4-vector and obey the continuity equation in virtue of the Dirac equation. It may be confirmed directly within the spinor formalism that the complex entity $(\Psi^T \gamma^2 \Psi, \Psi^T \gamma^2 \gamma^0 \gamma^i \Psi)$ is indeed a (gauge-dependent) conserved 4-vector. Combining these results with the similar ones for $\psi_I{}^2$ (for which $i \to -i$ in (78) and (79)), the Dirac current follows from the superposition of the two partial Majorana currents, as in (72) and (73).

### 7.2 Polar representation. The quantum potential

The second example we consider where the Dirac equation implies a conserved non-negative density is obtained by summing the equations for the functions $\psi_R{}^2$ and $\psi_I{}^2$. This yields a continuity equation involving the amplitude squared of the wavefunction:

$$\partial \sin\alpha|\psi|^2/\partial t + \partial_i(\sin\alpha\,|\psi|^2 v^i) + \partial_r(\sin\alpha|\psi|^2\,v^r) = 0 \tag{80}$$

where

$$v^i = (\psi_R{}^2 v_R^i + \psi_I{}^2 v_I^i)/|\psi|^2, \quad v^r = (\psi_R{}^2 \tilde{v}_R^r + \psi_I{}^2 \tilde{v}_I^r)/|\psi|^2 \tag{81}$$

are the local means of the partial velocities. Eq. (80) corresponds to the 'hydrodynamic' version of the field equation (38), obtained by using the polar field variables defined by the decomposition $\psi = |\psi|e^{iS/\hbar}$ in place of the functions $\psi_R, \psi_I$. In terms of the polar variables the velocities are

$$\left.\begin{array}{l} v^i = c(R_{1i} + \varepsilon_{ijk}R_{1j}\widehat{m}_k|\psi|/|\psi|), \\ v^r = -\delta_3^r \omega + 2c\left[A_i^r\left(\dfrac{\hat{n}_1 \partial_i|\psi|}{|\psi|} - \dfrac{\hat{n}_1 S \partial_i S}{\hbar^2}\right) - B_1^r\left(\dfrac{\widehat{m}_i|\psi|\partial_i|\psi|}{|\psi|^2} + \dfrac{\widehat{m}_i S \partial_i S}{\hbar^2}\right)\right]. \end{array}\right\} \tag{82}$$

In this case the density $\sin\alpha|\psi|^2$ and velocities are gauge independent. Note that, unlike the polar representation of the non-relativistic spin $\frac{1}{2}$ wave equation in the angular representation [3], the translational velocity $v^i$ is an angular rather than a translational gradient and its potential is $|\psi|$ rather than $S$ (the latter features as a potential for $v^r$). Using (31), the mean translational speed inherits the perpetual superluminality property (40) of its constituents: $v = \sqrt{v^i v^i} \geq c$.

The real relation implied by the Dirac equation (38) complementary to (80) is the Hamilton-Jacobi-like equation

$$\frac{\partial S}{\partial t} + v^i \partial_i S + v^r \partial_r S + Q = 0, \qquad Q = \frac{c\partial_i(|\psi|(\boldsymbol{R}_1 \times \widehat{\boldsymbol{m}}S)_i)}{|\psi|}. \tag{83}$$

A virtue of the polar angular formulation is that it demonstrates that there is a quantity implicit in the Dirac equation, which we denote $Q$, that may be identified as the 'quantum potential' for a massive relativistic spin $\frac{1}{2}$ system. This version of the quantum potential possesses the properties expected of it [3]: it depends on the form of $\psi$ rather than its absolute magnitude; it is second order in the configuration space coordinates; it is gauge invariant; and it appears as a kind of non-classical addition to the Hamilton-Jacobi equation if the kinetic energy is identified as the term $v^i \partial_i S + v^r \partial_r S$. Unusually, $Q$ depends on $S$ as well as $|\psi|$.

We now see why the polar representation is not apposite when seeking a trajectory construction of the wavefunction for which the equations are written just in terms of the density and velocities (as in methods A and B in Sect. 1). First, the angular velocity $v^r$ involves $S$ and cannot be expressed solely in terms of the density. Hence, the first-order trajectory equation $\dot{\theta}^r = v^r$ is not of the form (12). Second, the presence of $S$ in the quantum potential implies that we cannot obtain an alternative second-order formulation either. To see this, we use the material time derivative $\partial/\partial t|_{q_0,\theta_0} = \partial/\partial t|_{x,\alpha} + v^j \partial_j + v^r \partial_r$ and apply it to $v^r$ in (82). Replacing the time derivatives using (80) and (83), $\ddot{\theta}^r$ is found to depend on functions of derivatives of $S$ that cannot be reduced to functions of just $\dot{q}^i$, $\dot{\theta}^r$ and $\sin\alpha|\psi|^2$.

We conclude that, using the polar variables, it is not possible to obtain self-contained trajectory equations, in either the first- or second-order cases. Of course, trajectories $(q^i, \theta^r)$ can be derived from given translational and angular velocity fields, and these conserve the quantity $|\psi|^2 d^3x d\Omega$. This may provide an alternative causal interpretation of the Dirac equation, a possibility we discuss elsewhere. A noteworthy feature is that (83) states that the negative rate of change of the phase along a configuration space trajectory is the quantum potential: $-\partial S/\partial t|_{q_0,\theta_0} = Q$.

## 8 Effect of external potentials

We shall sketch how our results are modified by including an external 4-potential $(cA_0, A_i)$ in the Dirac equation, whose angular form becomes

$$i\hbar\, \partial\psi/\partial t - cA_0\psi = c\hat{n}_1\widehat{m}_i(i\hbar\partial_i\psi - A_i\psi) + imc^2\hat{n}_3\psi. \tag{84}$$

The equation can still be written in continuity form in $\mathfrak{M}$ but the external field renders this complex, which introduces a mutual coupling between $\psi_R$ and $\psi_I$. Thus, for $\psi_R$,

$$\partial \sin\alpha\, \psi_R/\partial t - \partial_i(c \sin\alpha\, \hat{n}_1 \widehat{m}_i \psi_R)$$
$$-2(c/\hbar)\partial_r[\sin\alpha\,(A_i^r(A_0 \widehat{m}_i \psi_I/3 + A_i \hat{n}_1 \psi_I) + mc\delta_3^r \psi_R)] = 0, \tag{85}$$

while for the similar equation for $\psi_I$ the potentials couple to $\psi_R$. In (85), we have replaced $A_0 \psi_I$ by $\widehat{m}_i \widehat{m}_i A_0 \psi_I/3$ ($\psi_I$ being an eigenstate of the total angular momentum) in order to include the $A_0$ term in a divergence. The form (39) for the translational velocity $v_R^i$ stays the same while it is the angular velocity that carries the coupling (and similarly for $\psi_I$). The two real continuity equations for $\psi_R$ and $\psi_I$ must now be solved simultaneously to obtain $\psi$. This may be achieved using our trajectory method by extending equation (12) to two equations and allowing their right-hand sides to depend on two densities. To solve for the two congruences $q_R^i$, $\theta_R^r$ and $q_I^i$, $\theta_I^r$, we use relation (49) to write the trajectory equations in terms of a common set of independent variables. This method of solving coupled continuity equations has been illustrated previously [8]. Finally, our analysis of the Lorentz covariance of the material picture can be applied to this case. Hence, in all respects, our constructive trajectory theory encompasses the external field case.

Passing to equation (76) for the non-negative density $\sin\alpha\, \psi_R^2$, it does not seem possible to incorporate the external field components in divergence terms so our trajectory technique of solving Dirac's equation does not apply to this case.

Finally, we consider the Dirac equation in the polar representation. The potentials are introduced by making the $\alpha$-independent replacement $S \to S + \int cA_0 dt + A_i dx^i$ in the continuity and Hamilton-Jacobi equations, (80) and (83), and in the velocities (82). All the resulting relations are invariant under an $x$-local gauge transformation $S' = S(x,\alpha,t) + f(x,t), A_0' = A_0 - \partial f/c\partial t, A_i' = A_i - \partial_i f$. In this context, the negative rate of change of the phase along a trajectory is the total potential energy: $-\partial S/\partial t|_{q_0,\theta_0} = cA_0 + v^i A_i + Q$.

## 9 Conclusion

In accordance with method B in Sect. 1, we have shown that the quantum state of a massive relativistic spin $\frac{1}{2}$ system may be represented exactly in material terms by two sets of trajectory and initial-density functions, $q_R^i(t), \theta_R^r(t), \psi_{R0}$ and $q_I^i(t), \theta_I^r(t), \psi_{I0}$. By varying the trajectory labels, each set ($R$ or $I$) describes an ensemble of 3-paths which, together with a mass-dependent operator produced by the evolution of the angular coordinates, generates the time dependence of a Majorana spinor ($\Phi_R^a$ or $i\Phi_I^a$; see (48)). Each set is governed by self-contained first-order equations (41) and (42) that are Lorentz covariant and globally phase dependent. The trajectory construction of the gauge invariant evolution of an arbitrary free Dirac spinor follows by superposition of the two so-constructed Majorana spinors, via formula (50). In the material picture a Lorentz transformation encompasses a relabelling substitution which has two consequences that are key in establishing covariance: a novel velocity transformation law, and a covariant notion of simultaneity. Developing the scheme, we showed how our constructive method extends to the inclusion of external potentials. We also showed how the future-causal Dirac 4-current may be expressed as the mean over the two superluminal translational flows, weighted with non-negative densities that each obey a continuity equation. The latter property provides an alternative trajectory construction of free wavefunction propagation. Finally, we examined the polar representation of the Dirac equation, which also implies a non-negative conserved density, and exhibits the Dirac analogue of the quantum potential. We explained why the polar version does not map into a stand-alone constructive trajectory theory (of the type A in Sect. 1).

Our methods A and B (Sect. 1) show that, if one desires to utilise trajectories in driving the evolution of $\psi_0$ into $\psi(t)$, the path integral is not the only option. Indeed, our constructions reflect the prevailing physical situation more closely, and are conceptually simpler, than Feynman's. Whereas Feynman employs the propagator and attributes equal amplitudes to 'all possible paths', in particular to those passing through regions where the wavefunction is small or even zero, our methods apply directly to the wavefunction: its value at each configuration-spacetime point is built from at most one (method A) or two (method B) paths, and the initial wavefunction contributes to the path dynamics. Moreover, in application to the Dirac equation, our method utilizes ordinary calculus and a fully Lorentz covariant propagation driven by ensembles of trajectories in 3-space, features that are difficult to achieve with the path integral [28].

Our primary concern has been to demonstrate the generality and consistency of the alternative trajectory conception of Dirac propagation. A possible drawback of the approach is the complexity of the trajectory equations, and its practicality as a computational tool is an open question. With regard to the physical meaning of the formalism, we have not offered an interpretation of the conserved quantities $\psi_R d^3x d\Omega$ and $\psi_I d^3x d\Omega$ whose associated variable-sign densities $\psi_R$ and $\psi_R$ are determined by Majorana amplitudes (or of the conserved quantities that involve the squares of these densities in the alternative constructive method of Sect. 7.1). The gauge-dependent character of the Majorana state variables, alongside the collective gauge invariance of their field equations, suggests that these entities play a role analogous to that of the electromagnetic potentials. The gauge dependence of the potentials does not preclude their providing a representation of the state of the electromagnetic field, and of course the gauge-invariant physical field is derived from them. In our material picture the 'gauge potentials' are the trajectories and initial densities, and the physical field, the equivalence class of spinor fields $\Psi^a$ connected by global phase transformations, is derived from them via formula (50) (this is connected to 'wavefunction potentials' [6]).

There are two technical issues that warrant further investigation. One is the possible appearance of singularities in the trajectory equation (41) in nodal regions (where $\psi = 0$). The second is a possible underdetermination in the theory. This stems from the leeway in the identifications we made in Sect. 5 of the two sets of translational and angular velocities as functions of the wavefunction. For example, the associated currents in $\mathfrak{M}$ are susceptible to the addition of divergence-free vectors, a freedom that appears to be compatible with the uniqueness property of the Dirac 4-current [18,19]. This issue may impinge on specific properties of the paths, such as superluminality and the possible occurrence of singularities, which could be artefacts of our identifications.